\documentclass[referee]{aa} 
\usepackage{graphicx}
\usepackage{longtable}
\usepackage{txfonts}
\begin{document}
\title{
Lunar occultations of 184 stellar sources in two crowded regions
towards the galactic bulge.
\thanks{Based on observations made with ESO telescopes at Paranal Observatory}
}

\titlerunning{
Lunar occultations of 184 stellar sources
}

   \author{A. Richichi\inst{1}\fnmsep\inst{2}
             \and
          W.P. Chen\inst{3}\fnmsep\inst{4}
          \and
          O. Fors\inst{5}\fnmsep\inst{6}
          \and
          P.F. Wang\inst{4}
          }

   \offprints{A. Richichi}

   \institute{
   National Astronomical Research Institute of Thailand,
   191 Siriphanich Bldg., Huay Kaew Rd., Suthep, Muang, 
Chiang Mai  50200, Thailand \email{andrea@narit.or.th}
\and
on leave from
             European Southern Observatory,
Karl-Schwarzschild-Str. 2, 85748 Garching bei M\"unchen, Germany
         \and
Graduate Institute of Astronomy, National Central University,
 300 Jhongda Road, Jhongli 32054, Taiwan
          \and
Department of Physics, National Central University,
 300 Jhongda Road, Jhongli 32054, Taiwan
         \and
Departament Astronomia i Meteorologia and Institut de Ci\`encies 
del Cosmos (ICC),
Universitat de Barcelona (UB/IEEC), 
Mart\'{\i} i Franqu\'es 1, 08028 Barcelona, Spain
         \and
             Observatori Fabra, Cam\'{\i} de l'Observatori s/n, 08035 
Barcelona, Spain
             }

  \abstract
{
Lunar occultations (LO) provide a unique combination
of high angular resolution and
sensitivity at near-infrared wavelenghts.
At the ESO Very Large Telescope, it is possible to
achieve about 1 milliarcsecond (mas) resolution
and detect sources as faint as K$\approx$12\,mag.
}
{
We have taken advantage of a passage of the Moon
over two crowded and reddened regions in the direction of
the inner
part of the galactic bulge, in order to obtain
a high number of occultation light curves over
two half nights. Our goal was to detect and
characterize new binary systems, and to investigate
highly extincted and relatively unknown infrared 
sources in search of circumstellar shells and
similar peculiarities. Our target list included
a significant number of very late-type stars, but
in fact the majority of the sources was without
spectral classification.
}
{
A LO event requires the sampling
of the light curve at millisecond rates in order to permit
a detailed study of the diffraction fringes. For this,
we used the so-called burst mode of the ISAAC instrument
at the Melipal telescope. Our observing efficiency was
ultimately limited by overheads for telescope pointing
and data storage, to about one event every three minutes.
}
{
We could record useful light curves for 184 sources.
Of these, 24 were found to be binaries or multiples,
all previously unknown. The
projected separations are as small as 7.5\,mas, and
the magnitude differences
as high as $\Delta$K=6.5\,mag.
Additionally we could establish, also for the
first time, the resolved
nature of at least two more stars, with
indication of circumstellar emission.
We could also put upper limits on the angular
size of about 165 unresolved stars, an 
information that combined with previous and
future observations will be very helpful in establishing
a list of reliable calibrators for long baseline
interferometers. 
}
{
Many of the newly detected companions are beyond the
present capabilities of other high angular resolution techniques,
however some could be followed up by long baseline
interferometry or adaptive optics.
From estimates of the stellar density
we conclude that, statistically, most of the companions
are not due to chance alignments.
}

   \keywords{
Techniques: high angular resolution --
Occultations --
Stars: binaries: general --
Stars: fundamental parameters --
Stars: circumstellar matter --
Infrared: stars
            }
   \maketitle

\section{Introduction}
Lunar occultations (LO) are simple, economic, time efficient, and they
provide
high angular resolution far exceeding the diffraction
limit of any single telescope. In fact the resolution of LO
is matched at present only by long-baseline interferometry (LBI), 
which however is far from being simple, economic or time efficient.
At a large telescope, LO also offer a superior sensitivity.
Clearly LO have significant limitations of their own, 
in particular the fact that they are fixed time events of
sources selected randomly by the apparent lunar motion.
Another important limitation is the fact that a single
LO event only provides information along the direction
of the limb motion, i.e. it is intrinsically one-dimensional.
This can be overcome if more events of the same source
are observed, either from different locations or in the course of
different lunar passages. Admittedly, this is difficult to
achieve if a large telescope is to be employed. 
On the other hand, it is worth mentioning a significant
bonus of the LO technique, namely the possibility to apply
statistical methods that yield the actual brightness profile
and not just a diameter or a binary separation. This makes
LO ideal to study sources with complex geometries.

Given the characteristics and limitations of LO, this method
is best suited for routine observations of large samples
of targets with the aim of
serendipitous detections of binary stars and resolved sources.
At a large telescope where observing time is precious,
there are essentially two ways to implement this strategy:
wait for favorable episodes in which the Moon covers
a large number of sources in a brief period of time (see
Richichi et al. \cite{PaperI}, \cite{PaperII}), or
obtain small chunks of observing time whenever they become
available (see Richichi et al. \cite{PaperIII}).
These three latter references (Paper~I to Paper~III,
respectively) include a detailed description of 
how the LO method has been implemented at the ESO
Very Large Telescope (VLT), both in terms
of instrumental solutions, of observational strategy, of data reduction,
and of performance. Thus, in the present paper we
will  linger only very brief on these aspects, and 
concentrate instead on the results and the conclusions.

\section{Observations and Data Reduction}\label{data}
Following the strategy of maximum return for minimum
observing time, we observed the passage of the Moon over
two crowded regions rich in near-infrared sources
during the two first half nights of September 25 and 26, 2009.
As in the previous papers we used ISAAC, this time
 at the Melipal telescope (UT3).
The instrument was operated in burst mode, reading out
a 32x32 pixels ($4\farcs7 \times 4\farcs7$) subwindow
with a time sampling of 3.2\,ms which was also the effective
integration time. We used mostly sequences of 5000 frames, or
16~s, per star. In a few cases we chose slightly shorter
or longer sequences to manage situations of occultations
very close to each other in time, or cases of two sources
occulted within the same field of view (see  Sect.~\ref{widebin}).
A broad band K-short filter was employed
for most of the sources, except for three bright stars for which
we used a narrow band filter centered around 2.07$ \mu$m
to avoid saturation. All events were disappearances, with a lunar 
phase of about 49\% and 59\% in the two nights. The seeing
was generally good with extremes of $ 0\farcs44 $ and
$ 1\farcs1 $, and we could follow the Moon to a maximum
airmass of 2.8. Over the two nights, we recorded 202 events in
about 9.5 hours, but a small fraction of data proved to
be not usable and in practice we have a total of  184 confirmed
 light curves. Using the magnitudes from the
2MASS Catalogue, 
the brightest and faintest star
had K=2.7 and 10.9\,mag respectively. A summary
of the main statistics of the sample is
given in Table~\ref{tab:stats}, while a
detailed
list is provided in Table~\ref{lo_complete} which 
follows the same style of Paper~III.
\begin{table}
\caption{Statistics of observations}
\label{tab:stats}
\centering          
\begin{tabular}{lcc}
\hline 
\hline 
\multicolumn{1}{c}{Night}&
\multicolumn{1}{c}{25 Sep, 2009}&
\multicolumn{1}{c}{26 Sep, 2009}\\
\hline 
Duration & 4$^{\rm h}$22$^{\rm m} $ & 5$^{\rm h}$03$^{\rm m} $  \\
Sources & 86 & 98 \\
Average RA & $18^{\rm h} 05^{\rm m}  $& $18^{\rm h} 58^{\rm m}  $\\
Average Dec & $ -25\degr 32\arcmin $& $-23\degr 48\arcmin $\\
Airmass min & 1.02 & 1.00 \\
Airmass max & 2.69 & 2.79 \\
Average seeing& $0\farcs67$ & $0\farcs72$ \\
K min (mag) & 2.74 & 4.60 \\
K max (mag) & 8.97 &10.86 \\
K median (mag) & 6.74 & 8.52 \\
Average J-K (mag) & 2.3 & 1.0 \\
Binaries \& Triples & 14 &10  \\
Extended sources & 2  & 0  \\
\hline 
\hline 
\end{tabular}
\end{table}

The process of extracting LO light curves from
the burst mode data cubes and the
corresponding analysis has
already been described, particularly
in Paper~I. As before, we used two versions
of the data analysis, both a
model-dependent and model-independent one
(ALOR and CAL respectively,
Richichi et al. \cite{richichi96}
and
Richichi \cite{CAL}).
Upper limits on the angular sizes of unresolved
sources were computed using 
the approach described in
Richichi et al. (\cite{richichi96}).
One significant innovation in our arsenal of
data reduction tools is a method to deal
with light curves which appear to exhibit a
variable lunar rate, i.e. for which presumably
the local limb slope is changing across the
duration of the event. This effect is generally quite small,
e.g. 1$ \degr $ or 2$ \degr $ over a few
tenths of a second, but given the very high
signal-to-noise ratio (SNR) of some of our data it is
occasionally noticeable. This method, as well
as other considerations on lunar limb effects,
is part of a separate paper in preparation.

\section{Results}\label{results}
\begin{figure}
\includegraphics[angle=-90,width=8.8cm]{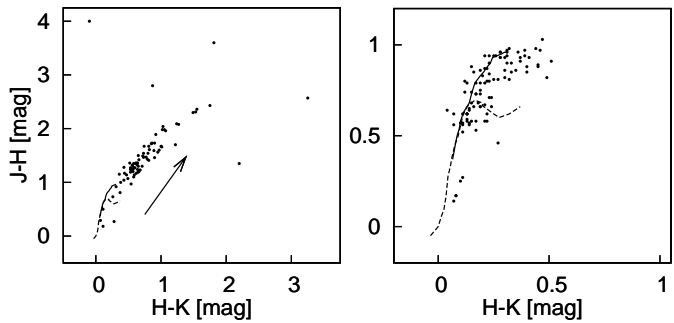}
\caption {Color-color diagram for the sources (dots)
for which we recorded the LO events 
listed in Table~\ref{lo_complete}.
The left and right panels are for the nights of
September 25 and 26, respectively. Note the different
scales.
The solid and dashed lines are the loci of the unreddened giant and
dwarf stars, respectively, according to 
Bessell \& Brett (\cite{bessellbrett}).
The arrow is the extinction vector for A$_{\rm V}$=10\,mag, according to
Rieke \& Lebofsky (\cite{riekeleb}).
}
\label{fig_km}
\end{figure}

The stars of the first night were clustered in a much
more interior region of the bulge than those of the
second night. 
The minimum angular distances from the
Galactic Center were just under 5$\degr$ and almost
17$\degr$ in the two cases. It is then no surprise that the stars
occulted in the first night were in general much 
redder, as shown in Fig.~\ref{fig_km}. It can be seen
that in this case the reddening can be well explained by general
interstellar extinction, with A$ _{\rm V} $ of up
to more than 15\,mag. This is consistent with
the map of general extinction in this area derived
from dust infrared emission by Schlegel et al. 
(\cite{1998ApJ...500..525S}) which show levels of
$6 \la {\rm A}_{\rm V} \la 30$ for our sources
of the first night.
On the second night, little
or no reddening is observed and the average (J-K) colors
are in line with the typical value of field stars,
see Table~\ref{tab:stats}. The table also shows
that on the first night the occulted stars were on
average significantly brighter than those of the
second night: this is the result of the first region
being much more crowded, and therefore giving us
the opportunity to select occultations of brighter stars in the
typical interval of 3 minutes. As mentioned in our
 previous papers, the number of occultation events
 to the sensitivity limit of K$\approx 12$\,mag
far exceeds the practical possibilities of data recording,
and we prioritized the events based on
K brightness, near-IR colors, and source type and
literature entries when available.
We note that there is no overlap with the sources included in Papers~I
to III.

The majority of the targets have
near-IR colors  consistent with those of giant or dwarf stars
after making proper adjustments for extinction,
but in few cases more peculiar colors are also observed. 
We will comment on these,
as well as on other cases where the reddening might have
circumstellar origins, later in this section.
Due also to the reddening, the stars in our sample
are generally faint or even undetected at visual wavelengths and 
are thus relatively poorly studied. Only about 7\% have
a spectral classification and about 16\% have 
an entry
in the Simbad database. Even for these, the 
available literature
is generally very scarce.

In Table~\ref{tab:results} 
we list the sources for which we found a positive result, 
either as a binary or multiple star or
with a resolved angular diameter or with
circumstellar emission. For ease
of reference, the table is
by right ascension and not by time of occultation.
It follows a similar format and the same
conventions of our previous papers.

We discuss first
the sources which were found to be resolved and
for which we can provide some context from existing
literature, as well as those for
which literature entries exist but which we
found to be unresolved. We then discuss briefly those without
any known cross-identification, 
and a few
relatively wide pairs.
Finally, we provide
some considerations and general results on the 
statistics of detected binaries and on the
unresolved sources. We do not
show all the light curves
and their fits. Instead, we provide one
example each for a binary, an unresolved and an
extended source.

\addtocounter{table}{1}
\begin{table*}
\caption{Summary of results.}
\label{tab:results}
\centering          
\begin{tabular}{lcrrrrrrrl}
\hline 
\hline 
\multicolumn{1}{c}{(1)}&
\multicolumn{1}{c}{(2)}&
\multicolumn{1}{c}{(3)}&
\multicolumn{1}{c}{(4)}&
\multicolumn{1}{c}{(5)}&
\multicolumn{1}{c}{(6)}&
\multicolumn{1}{c}{(7)}&
\multicolumn{1}{c}{(8)}&
\multicolumn{1}{c}{(9)}&
\multicolumn{1}{c}{(10)}\\
\multicolumn{1}{c}{Source}&
\multicolumn{1}{c}{V (m/ms)}&
\multicolumn{1}{c}{V/V$_{\rm{t}}$--1}&
\multicolumn{1}{c}{$\psi $($\degr$)}&
\multicolumn{1}{c}{PA($\degr$)}&
\multicolumn{1}{c}{CA($\degr$)}&
\multicolumn{1}{c}{SNR}&
\multicolumn{1}{c}{Sep. (mas)}&
\multicolumn{1}{c}{Br. Ratio}&
\multicolumn{1}{c}{Comments)}\\
\hline 
18022636-2555373 & 0.2599 &  $-$10.7\% & $-$3.3 & 133 & 58 & 119.1 & 9.9 $\pm$ 0.2 & 25.0 $\pm$ 0.1 & K$_1$=6.0, K$_2$=9.5 \\
18023182-2539066 & 0.5331 &  $-$9.8\% & $-$18.0 & 48 & $-$27 & 207.2 & 22.1 $\pm$ 0.2 & 152.2 $\pm$ 1.6 &   K$_1$=5.2, K$_2$=10.7 \\
18024042-2549567 & 0.4613 &  $-$7.2\% & $-$5.7 & 104 & 28 & 60.6 & 40.2 $\pm$ 0.5 & 58.8 $\pm$ 1.5 &   K$_1$=7.2, K$_2$=11.6 \\
18025652-2532392 & 0.5723 &  3.1\% & 3.7 & 51 & $-$23 & 160.4 & 20.5 $\pm$ 0.2 & 76.6 $\pm$ 0.8 &   A-B: K=5.6, 10.3 \\
18025652-2532392 & 0.5723 &  3.1\% & 3.7 & 51 & $-$23 & 160.4 & 127.7 $\pm$ 0.3 & 173.2 $\pm$ 2.0 &   A-C: K=5.6, 11.2 \\
18032656-2526103 & 0.4677 &  2.8\% & 1.5 & 28 & $-$46 & 98.2 & 17.0 $\pm$ 0.5 & 98.5 $\pm$ 2.1 &    K$_1$=7.2, K$_2$=12.7 \\
18033522-2523185 & 0.2700 &  $-$20.3\% & $-$6.5 & 6 & $-$67 & 108.2 & 7.3 $\pm$ 0.2 & 57.2 $\pm$ 0.9 &    K$_1$=6.4, K$_2$=10.8 \\
18040356-2523142 & 0.3208 &  $-$31.3\% & $-$14.6 & 9 & $-$63 & 70.7 & 7.8 $\pm$ 0.1 & 16.1 $\pm$ 0.1 &    A-B: K=7.0, 10.0 \\
18040356-2523142 & 0.3208 &  $-$31.3\% & $-$14.6 & 9 & $-$63 & 70.7 & 19.3 $\pm$ 0.3 & 45.8 $\pm$ 0.7 &    A-C: K=7.0, 11.2 \\
18040728-2539532 & 0.6170 &  $-$2.4\% & $-$3.5 & 88 & 15 & 149.4 & 142.2 $\pm$ 0.8 & 418.1 $\pm$ 15.3 &    K$_1$=4.9, K$_2$=11.4 \\
18041209-2544257 & 0.4742 &  $-$9.7\% & $-$6.4 & 106 & 33 & 17.8 &    &    & Extended $\approx$4.31\,mas \\
18050733-2543282 & 0.4660 &  14.2\% & 5.7 & 134 & 62 & 71.5 & 16.6 $\pm$ 0.6 & 91.5 $\pm$ 3.0 &    K$_1$=6.7, K$_2$=11.7 \\
18050977-2543351 & 0.5337 &  37.2\% & 14.6 & 145 & 73 & 6.2 & 33.7 $\pm$ 1.4 & 9.5 $\pm$ 0.6 &    K$_1$=7.4, K$_2$=9.9\\
18053438-2540309 & 0.4583 &  $-$4.9\% & $-$2.3 & 120 & 48 & 48.6 & 8.3 $\pm$ 0.3 & 27.7 $\pm$ 0.5 &    K$_1$=6.9, K$_2$=10.5 \\
18054372-2514234 & 0.4278 &  $-$0.5\% & $-$0.2 & 15 & $-$57 & 89.9 & 10.9 $\pm$ 0.2 & 46.3 $\pm$ 0.5 &    A-B: K=5.6, 10.3\\
18054372-2514234 & 0.4278 &  $-$0.5\% & $-$0.2 & 15 & $-$57 & 89.9 & 24.2 $\pm$ 0.2 & 62.0 $\pm$ 0.6 &    A-C: K=5.6, 11.2 \\
18054490-2540399 & 0.4026 &  $-$6.6\% & $-$2.5 & 126 & 54 & 99.6 & 10.5 $\pm$ 0.1 & 12.6 $\pm$ 0.0 &    K$_1$=3.2, K$_2$=6.0 \\
18063774-2529277 & 0.6971 &  $-$7\% & $-$8.8 & 84 & 12 & 6.2 &    &    & 
Extended $\approx$12.7\,mas\\
18073464-2534117 & 0.3753 &  8.1\% & 2.0 & 141 & 69 & 70.6 & 37.1 $\pm$ 0.2 & 20.2 $\pm$ 0.1 &   K$_1$=5.9, K$_2$=9.1 \\
18545320-2357035 & 0.4844 &  14.5\% & 10.3 & 37 & $-$32 & 45.6 & 15.1 $\pm$ 0.3 & 34.2 $\pm$ 0.7 &    A-B: K=7.9, 11.7 \\
18545320-2357035 & 0.4844 &  14.5\% & 10.3 & 37 & $-$32 & 45.6 & 12.7 $\pm$ 0.5 & 54.8 $\pm$ 1.8 &    A-C: K=7.9, 12.2 \\
18552223-2412461 & 0.5265 &  3.6\% & 4.4 & 101 & 31 & 12.7 & 43.7 $\pm$ 2.9 & 10.5 $\pm$ 0.2 &    K$_1$=9.8, K$_2$=12.3 \\
18552401-2349283 & 0.1628 &  $-$19.8\% & $-$3.9 & 352 & $-$74 & 41.3 & 16.6 $\pm$ 0.4 & 14.5 $\pm$ 0.1 &    K$_1$=7.6, K$_2$=10.5 \\
18553233-2412387 & 0.4008 &  $-$19.0\% & $-$15 & 85 & 16 & 89.2 & 8.3 $\pm$ 0.2 & 10.4 $\pm$ 0.1 &    K$_1$=6.5, K$_2$=9.9 \\
18565536-2344580 & 0.4824 &  $-$4.1\% & $-$2.7 & 22 & $-$42 & 51.4 & 13.0 $\pm$ 0.8 & 44.8 $\pm$ 1.2 &    K$_1$=7.8, K$_2$=11.9 \\
18580265-2337079 & 0.3949 &  $-$24.3\% & $-$13 & 9 & $-$57 & 19.9 & 35.1 $\pm$ 2.1 & 32.8 $\pm$ 1.3 &    K$_1$=8.5, K$_2$=12.3 \\
19001313-2340112 & 0.7733 &  0.7\% & 1.3 & 85 & 19 & 95.0 & 17.8 $\pm$ 0.4 & 86.9 $\pm$ 2.1 &    K$_1$=6.3, K$_2$=11.1 \\
19001602-2349574 & 0.2667 &  $-$16.9\% & $-$4.1 & 129 & 63 & 29.8 & 13.8 $\pm$ 0.5 & 11.7 $\pm$ 0.1 &    K$_1$=7.4, K$_2$=10.0 \\
19010885-2317367 & 0.4861 &  $-$16.3\% & $-$7.5 & 11 & -58 & 7.3 & 129.6 $\pm$ 1.4 & 1.30 $\pm$ 0.01 &    K$_1$=10.4, K$_2$=10.6 \\
19011187-2323074 & 0.8207 &  1.7\% & 2.3 & 46 & $-$22 & 10.0 & 34.4 $\pm$ 3.3 & 13.7 $\pm$ 0.7 &    K$_1$=9.6, K$_2$=12.4 \\
\hline 
\hline 
\end{tabular}
\end{table*}

\subsection{Sources with known cross-identifications}\label{resolved}

\object{18022636-2555373}
is {\object{HD 314951}, a V=9.8\,mag star without
bibliographical entries.
Its colors suggest either a K5III or a slightly reddened K5V,
placing this star in the foreground of our
observed region. 
Fig.~\ref{fig_binary}
illustrates our detection
both by a model-dependent and a model-independent
approach, as used for all the binary stars in this paper.

\begin{figure}
\includegraphics[angle=-90,width=8.8cm]{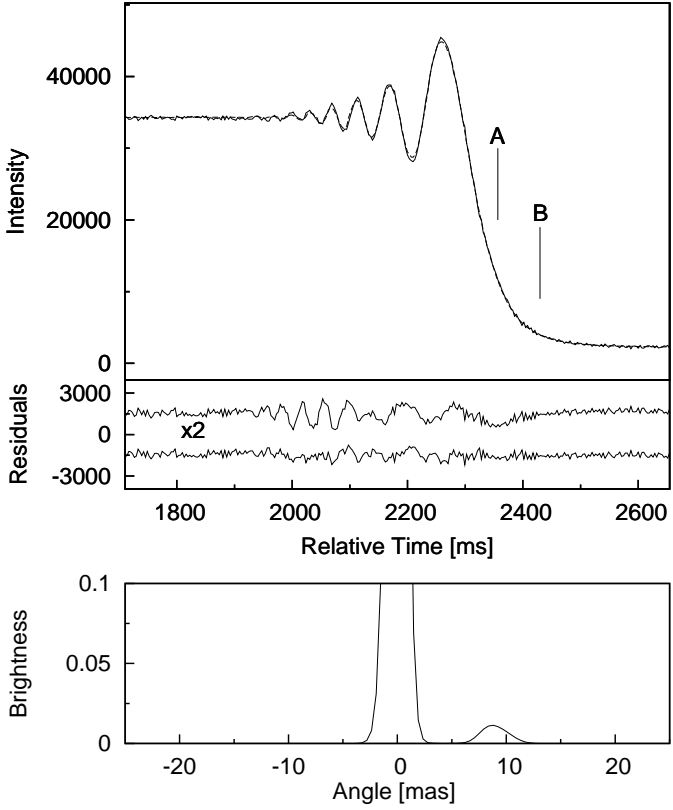}
\caption {{\it Top:} upper panel, data (solid) and best fit (dashed line)
for
{\object{18022636-2555373}}.  The lower panel shows,
displaced by arbitrary offsets and enlarged
for clarity, the residuals of the fits by a point-like 
(above)
and  by a binary star model including corrections due
to slope variations (see text).
The normalized $\chi^2$ values are 4.42 and 1.46, respectively.
The times of the geometrical occultation
of two stars are also marked, with their difference
corresponding to a separation of 9.9\,mas.
{\it Bottom:} brightness profile reconstructed by the
model-independent CAL method.
}
\label{fig_binary}
\end{figure}

\object{18041207-2544225} is part of a $3\farcs9$ wide pair, as detailed in 
Sect.~\ref{widebin}.
The star to the south, \object{18041209-2544257}, is brighter, with 
K=5.59\,mag, and unresolved with an
upper limit of $0.50\pm0.25$\,mas.  It has reliable 2MASS photometric 
measurements in all JHK bands,
and a position coincident with a WISE and a Spitzer detection, with 
[5.8]=1310\,mJy and [8.0]=850\,mJy.

The northern star, \object{18041207-2544225}, appears to be either 
marginally resolved with an angular diameter of 4.3\,mas or, 
more likely, associated with faint extended emission. Its 2MASS fluxes are 
uncertain, because of its proximity to the bright star to the south.  
Its Spitzer fluxes [5.8]=30\,mJy and [8.0]=15\,mJy are also unreliable. IRAS 
detected only one source, \object{18010-2544} with [12]=1178\,mJy.  
If this flux
came mostly from the southern source, then this star would have
a mid-IR excess. We stress that the SNR 
of our LO observation of the northern component is low, $\sim 18$. 
Further observations 
including spatially resolved near- and mid-IR photometry are needed to 
confirm and diagnose the nature of the extended emission.

\object{18044848-2536204} is \object{HD 164972},
an F2-F3II  or F3Ib star (De Medeiros et al. \cite{med02}).
Its parallax of $2.28\pm0.89$\,mas and
the absence of significant reddening show it is in the
foreground of our sample.

\object{18054372-2514234}
is the known bright 
near-infrared source \object{IRC -30354}.
Its 2MASS colors, J-H=1.35 and H-K=0.62, 
and V magnitude of 13.5
are consistent with a highly
reddened late-type dwarf
(Hansen \& Blanco \cite{han75}).
The star exhibits a broad H$_\alpha$
line in emission (Velghe \cite{vel57}).
We detect two companions, 
with small projected separations that
place the system at the limit of the
resolution for adaptive optics (AO) on a very
large telescope, and similarly the
large magnitude differences make it
challenging for LBI.

\object{18060176-2526099} is \object{IRC -30355}, a
bright infrared source.
Its 2MASS colors are consistent with a late M giant, consistent with 
its spectral type of M6 determined by Hansen \& Blanco (\cite{han75}).  
Detected by AKARI, WISE, IRAS, Spitzer and MSX, it has a rising 
spectral energy distribution with prominent far-infrared excess.
Our measurement shows an unresolved source in K band with
an upper limit of 0.40\,mas.

\object{18063774-2529277}
is \object{IRAS 18035-2529}, an OH-IR source
detected in the 1612 MHz OH maser line, with a typical 
double-peaked profile 
(te Lintel Hekkert \cite{lin91}). It is also a 
prominent SiO maser source (Deguchi et al. \cite{dug04}). 
We note that this star has extreme colors
J-H=2.57 and H-K=3.25 but  in fact the
2MASS J-band value is unreliable and the J-H could
be even higher.
At K=9.0\,mag this source could not provide a very
high SNR by LO. Actually, our detected flux was
rather consistent with K=10.5\,mag, indicating
a significant variability which is not unexpected
in this kind of source. Both model-dependent and
model-independent fits to the data
point to a resolved source. In the first case,
a formal diameter of $12.7\pm0.5$\,mas is derived.
The model-independent analysis (Fig.~\ref{fig_extended})
provides however a better normalized $\chi^2$
and is also more consistent with the OH-IR
scenario, showing a a central star with two
asymmetric emission peaks at about 10-20\,mas
which we attribute to the inner rim of a
circumstellar shell.

\object{18545320-2357035}
was  surveyed by 
Hipparcos and included as \object{TYC 6860-1491-1}
in the re-analysis which led to the Tycho Double Star Catalogue 
 (Fabricius et al. \cite{Fabricius2002}). 
That the multiplicity was not detected by Hipparcos is
consistent with the its quoted performance.
With V=10.6\,mag and colors which do not show
evidence of significant reddening, this star is
probably in the foreground of our sample.
We also mention a fainter, relative distant
companion 
(see Sect.~\ref{widebin}).
This was not detected by Hipparcos due
to its faintness.

\object{18550997-2401055} 
is \object{HD 175159}, a star for which a LO was
previously reported by Evans \& Edwards (\cite{1983AJ.....88.1845E}).
As in our case, the star was  found to be single by these authors.

\object{18570946-2352076} and 
\object{18570922-2351582} are 
\object{CD-24 14860B} and 
\object{HD 175601} respectively, and form
a wide visual double system with separations ranging 
from $10\farcs3$ (epoch 1910) to $9\farcs9$ (epoch 1998) 
(Dommanget \& Nys \cite{2002yCat.1274....0D}, 
Mason et al. \cite{2001AJ....122.3466M}). 
Both stars were found unresolved by us.

\object{18580265-2337079} is 
\object{PPM 734659}, a V=9.9\,mag star for
which no literature entries are available.
From the relatively bright visual counterpart, 
the unreddened colors and the high proper motion,
 we conclude that the star is in the foreground
and 
the 
$\Delta$K=3.8\,mag companion
that we detect 
deserves further investigations for possible
orbit determinations.

\object{18594214-2342263} and \object{18594214-2342263}
are \object{AR Sgr} and \object{HD 176364}, respectively.
The former is a G4 star of RV Tau type classified as
a very likely post-AGB object by Szczerba et al. 
(\cite{2007AA...469..799S}), while
the latter is a G8IV star. In both cases
our non-binarity detection confirms previous results from LO 
photoelectric observations 
(Evans \& Edwards \cite{1983AJ.....88.1845E}, 
Radick et al. \cite{1984AJ.....89..151R}).

\begin{figure}
\includegraphics[angle=-90,width=8.8cm]{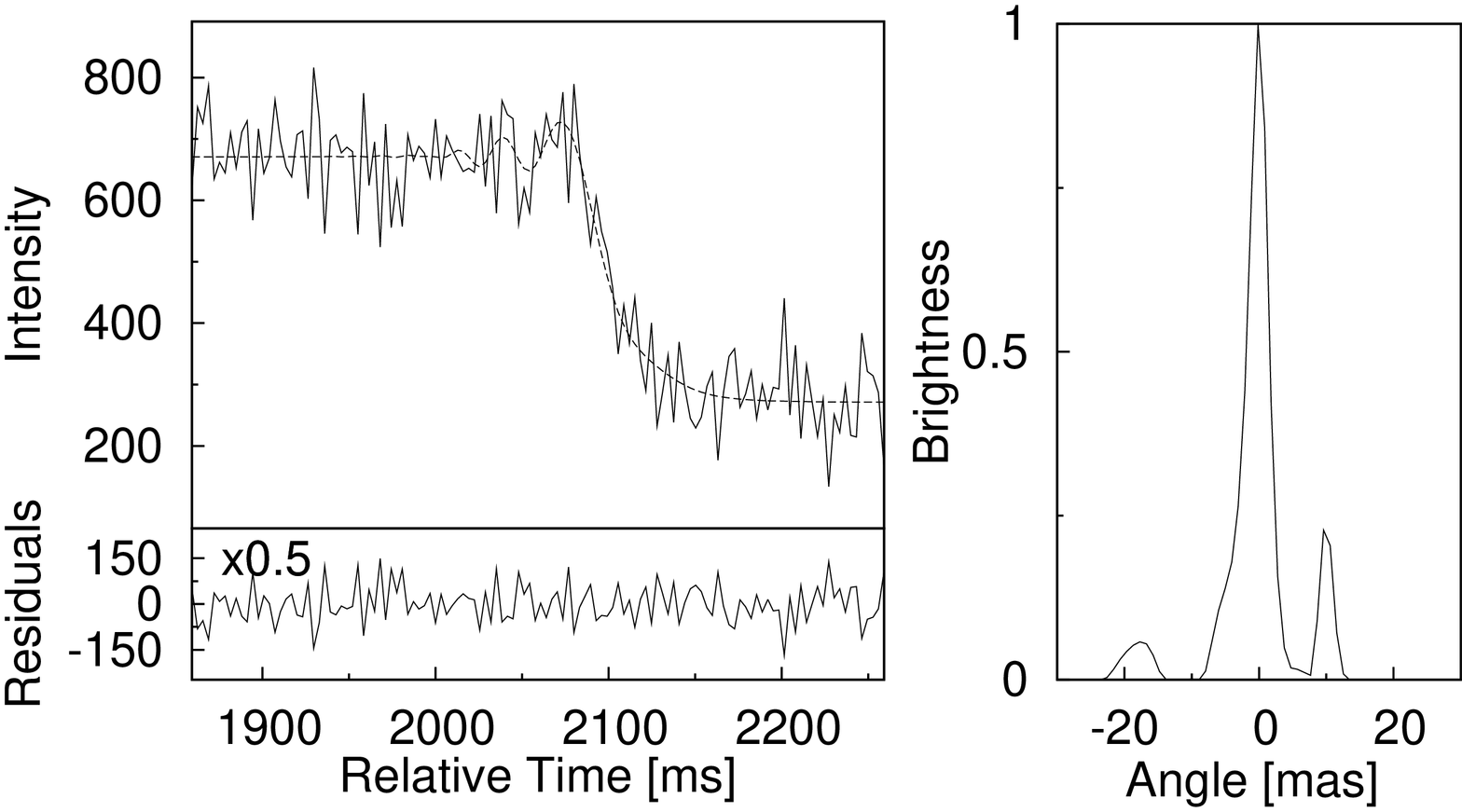}
\caption {
{\it Left:} upper panel, data (solid) and best fit (dashed line)
for
{\object{18063774-2529277}}.  The lower panel shows
the residuals of the fit by the profile shown to the right.
The normalized $\chi^2$ is 1.21. 
{\it Right:} brightness profile reconstructed by the
model-independent CAL method.
}
\label{fig_extended}
\end{figure}

\object{18071398-2516474} is \object{HD 315178},
a known B-type star with a strong H-alpha line in emission
(Merrill \& Burwell \cite{mer50}, Velghe \cite{vel57}).
From its 2MASS color, it obviously has an infrared
excess and it was detected by MSX at 8\,$\mu$m 
(Clarke et al. \cite{cla05}). From our measurement
we determine an upper limit on its K-band angular
size of 1.95\,mas.

\object{18072725-2506165} is
\object{HD 165517}, the optical counterpart of the 
X-ray source \object{1RXS J180725.9-250607}. 
This O9.5:Ibe emission-line star 
has been classified also 
as B0:Ia by Johnston et al.~(\cite{1994AA...289..763J}).
With an estimated distance of 
 2.1\,kpc (Kozok~\cite{1985AAS...62....7K}) it is in the foreground
 of our sample.
We do not detect any circumstellar emission and the
source appears unresolved with an upper limit 
of 1.2\,mas.

\object{18073464-2534117}
is \object{HD 165530}, a star with V=8.3\,mag and
spectral type G7III which has colors  consistent
with those of a giant without any
significant reddening. Indeed, the Hipparcos parallax
places it at 340\,pc albeit with a large error
(\object{HIP 88798}). The fact that our newly detected
companion was not revealed by 
Hipparcos is consistent with the performance
limits of this latter (Mignard et al. \cite{Mignard95}). 
We note that an inconsistency was found in the object type and literature
 references that the Simbad database associates to \object{HD 165530}. 
This is erroneously classified as a star in double system with a components
 separation of about $13{\farcs}3$ and linked to Gahm et al. (\cite{Gahm83}).
By double checking with archive images from several surveys, we noticed that
 the former reference is actually for \object{HD 165530B}, which 
effectively corresponds to the well-known \object{WDS J18089-2528} 
double system with the above mentioned components separation.

A few more sources in our sample  also have Simbad cross identifications,
however without bibliographical entries of any relevance.

\subsection{Resolved sources without
 known cross-identifications}\label{resolved-no}
The remaining entries in Table~\ref{tab:results}
have no known cross-identification in the Simbad
database. They include an additional 17 new binaries and
2 new triples, with
projected separations spanning the range
7-140\,mas.
Based on 2MASS, the primaries have K magnitudes
ranging from 3.2 to 10.4, and the companions
from 6.0 to 12.4. 
Thus, this sample of binaries represents
a good hunting ground for follow-up studies.
Although some of the sources might prove
beyond the possibilities of present techniques,
in many cases either AO or LBI should be able
to detect and study the systems.
The distribution in terms of projected separation
and magnitude differences is shown in Fig.~\ref{fig_distribution}.

\begin{figure}
\includegraphics[angle=-90,width=8.8cm]{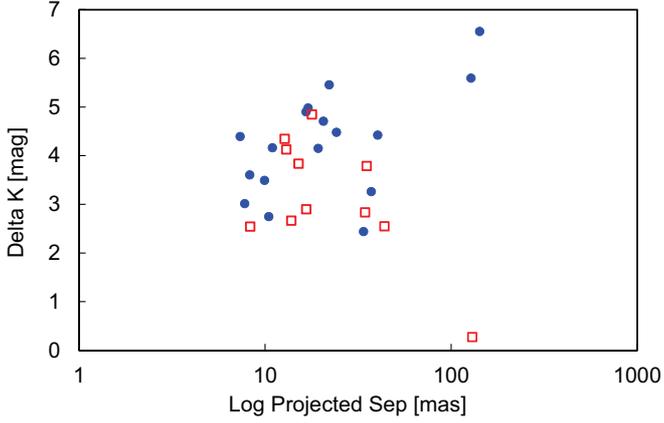}
\caption {
Distribution of the secondary and tertiary companions
in our sample, with respect to their respective primaries. 
The filled circles and empty squares represent
 the sources observed in the nights of Sep 25 and 26,
 respectively. Note that our  analysis did
 not extend beyond about $0\farcs15$, and that 
 the wide pairs of Table~\ref{table:wide}
 are not included.
}
\label{fig_distribution}
\end{figure}

We mention that \object{19010885-2317367}
is of some interest: it has components
of almost equal brightness and with a
projected separation of $0\farcs13$ it should
be quickly resolved with a variety of
methods. Given that the source
appears rather unreddened (J-H=0.27, H-K=0.11)
it is probably a not too distant physical pair.

The large distances to many of our sources, 
as implied by their substantial interstellar reddening,
and the relative crowding of the two regions
open the question of whether the binaries
that we report are physical pairs or chance
alignments. We have computed the average 
cumulative stellar density as a function of
magnitude in the two areas. In the most
crowded one, this amounts to 6.9$\times 10^{-4}$
and 3.0$\times 10^{-3}$ stars with K$\le$10
and $\le$12\,mag respectively, per square
arcsecond. The ISAAC subwindow size is
about 22~sq.~arcsec, but our effective
field of view is much smaller. In the direction
of the lunar motion we only analyze data
equivalent to few $0\farcs1$ on the sky,
since larger separations can be detected
by conventional imaging. In the perpendicular
direction, we are limited by the extraction
mask which is typically less than
half of the subwindow.
Although these quantities vary for each
event, our effective field of view is
in practice $\la$1~sq.~arcsec.
Therefore, we do not expect
any significative fraction of chance
associations in our sample.
The probability of
finding companions such as those listed
in Table~\ref{tab:results} is on
average 0.1\% in the first
night, and even less in the second night.

\subsection{Wide binaries}\label{widebin}
In six cases, our acquisition images showed two
sources present in the same 
$4\farcs7 \times 4\farcs7$ field of view.
By proper positioning in the subwindow and adjustment
of the number of frames we could record occultation
events for both.
They were all observed in the broad K$_{\rm}$ filter.
These pairs
are unresolved by 2MASS and therefore only one set of
J,H,K magnitudes is available, with the exception of the pair
\object{18041209-2544257} and \object{18041207-2544225}
already discussed in Sect.~\ref{resolved} for which however
the 2MASS magnitudes of the fainter star are unreliable.
From our LO analysis we could derive an accurate 
brightness ratio in the K-band for all these pairs, from
which we deduce the individual K magnitudes using
the 2MASS total value. For the above 2MASS-resolved
pair, we adopted the 2MASS magnitude for the brighter
source and derived that of the fainter one from its counts.

To compute separation and position angle (PA) values an average image was created from the occultation frames, and two-dimensional Gaussian PSF fitting (Bertin \& Arnouts \cite{bertin1996}) was used to find precise positions for the binary components. By comparing values obtained with different
 centroiding algorithms, we estimate accuracies of better than $0\farcs05$ 
 and $1\degr$ for separation and PA, respectively.
  The parameters for
  these wide pairs are listed in Table~\ref{table:wide}.
The suffixes 1 and 2 are in order of disappearance and
not of brightness. 
In Table~\ref{lo_complete} we can report only the total
J and H magnitudes for each pair.
Note that some of the individuals components of these
wide pairs were additionally found to be resolved, binary
or triple as already reported above.

\begin{table}
\caption{Parameters for the wide pairs}
\label{table:wide}
\centering          
\begin{tabular}{lrrrr}
\hline 
\hline 
\multicolumn{1}{c}{2MASS}&
\multicolumn{1}{c}{Sep ($\arcsec$)}&
\multicolumn{1}{c}{PA ($\degr$)}&
\multicolumn{1}{c}{K$_1$ (mag)}&
\multicolumn{1}{c}{K$_2$ (mag)}\\
\hline 
18030887-2535164 & 1.83 & 359 & 7.13 & 8.28 \\
18041207-2544225 & 3.94 & 354 & 9.41 & 5.59 \\
18050977-2543351 & 2.27 & 35 & 7.36 & 11.30 \\
18062439-2525353 & 2.30 & 208 & 10.14 & 7.33 \\
18545320-2357035 & 1.59 & 20 & 7.89 & 11.07 \\
18580265-2337079 & 2.80 & 280 & 9.37 & 9.06 \\
\hline 
\hline 
\end{tabular}
\end{table}

\subsection{Unresolved sources, variability and
 performance}\label{sect_unres}
As in previous papers, we used the
$\chi^2$-based 
procedure described in Richichi et al. (\cite{richichi96})
to assess whether a source was unresolved, and
determine an upper limit on its angular size.
These upper limits vary with SNR, being on average
2.2\,mas and reaching 0.25\,mas in the best case.
This was computed for 165 sources, but we note that
we are not including in this
count the individual components of binaries and multiples.

An example of a source established to be unresolved
is given in Fig.~\ref{fig_unresolved}, for 
\object{18012777-2538123} observed under average
seeing conditions in the first night. This source
has no known cross-identifications or literature
entries. With K=$6.6$ it falls close to the
median brightness of our sample, and with
SNR=110 it is representative of a good data quality although
by no means exceptional: our sample includes
many cases of better SNR, up to about 300.
The (J-K)=1.8 is consistent with the average
color of the sample and suggests no intrinsic
reddening. Fig.~\ref{fig_unresolved} shows
the fit with a point-like source, with a
normalized $\chi^2=1.013$. We compute an upper
limit of 0.75$\pm$0.30\,mas. For comparison,
a fit with a 2\,mas model increases the
normalized $\chi^2$ by 12.5\%.

\begin{figure}
\includegraphics[angle=-90,width=8.8cm]{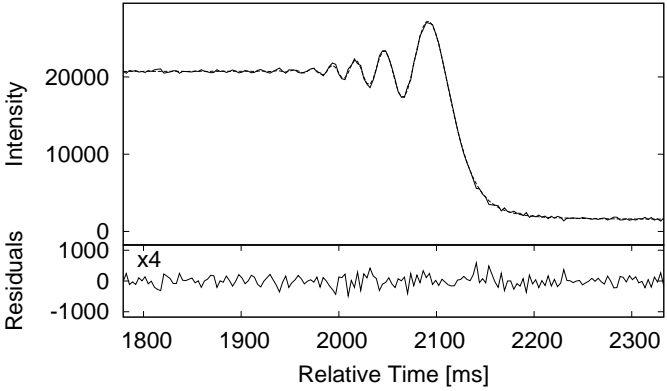}
\caption {
Same as the left side of Fig.~\ref{fig_extended}, for
\object{18012777-2538123} with an upper limit on the angular size 
of 0.75\,mas. The residuals are enlarged and
the normalized $\chi^2$ is 1.01. More details in Sect.~\ref{sect_unres}.
}
\label{fig_unresolved}
\end{figure}

 So far we have adopted the 2MASS K magnitudes
 at face value,
but we should now comment that these might
occasionally be inaccurate in two respects.
Firstly, by converting our observed counts
to a K magnitude using the ISAAC specific
response, we notice that about 9\% 
of the sources are up to 1\,mag brighter
than the 2MASS values, and about 17\% are
at least 0.3\,mag and up to almost 2\,mag fainter. 
We assume here
that up to 0.3\,mag dimming could be due
to airmass and our masking process.
Thus
variability is an important factor,
a fact perhaps not surprising when we
consider that most of the sources are
probably late-type giants. Indeed, the
most extreme case of flux deficit is that
of \object{18063774-2529277} already commented in Sect.~\ref{resolved}.
Secondly, some of the 2MASS magnitudes
are flagged as unreliable, due for example
to the presence of a nearby bright star. This applies
to three of the most extreme
points in Fig.~\ref{fig_km}, namely
\object{18063720-2514416} with J-H=4.00 and H-K=-0.10,
\object{18043132-2527356} (2.80, 0.87) and
\object{18024090-2552007} (1.35, 2.20).

In Papers~I to III we have already reported
details on the performance of sensitivity 
and angular resolution of the LO method
at the ESO VLT. Without further figures
we state here that also for the present
data we confirm an angular resolution
ranging from $\approx 0.5$\,mas for SNR$>$100
to $\approx 5$\,mas for SNR$\approx 10$.
These SNR values are typically achieved
for K magnitudes of $\approx 5$ and
$\approx$10, respectively. A comparison
with the early predictions on LO performance
at a very large telescope using subarrays
(Richichi \cite{richichi1994}) shows that
a significant improvement could still be
gained in sensitivity, by using instruments
with a pixel size matching the seeing
disk.

We also note that, adding to the previous
statistics reported in Papers~I to III, we
have now a database of about 350 unresolved stars
with reliable upper limits as small as 0.25\,mas,
determined in
a consistent fashion using LO at the VLT.
The range of magnitudes and of sizes
represents an ideal catalog of calibrators
for near-IR LBI, to which we can add the
substantial number of resolved and binary
stars which were not previously known and
that can be likewise flagged as non suitable
calibrators.
This database is still increasing thanks
to further observations currently under
analysis, and we plan to publish it
in the near future.

\section{Conclusions}
Using the LO technique with the ISAAC instrument
at the ESO VLT, we have carried out milliarcsecond
 resolution observations of 184 sources in two
 crowded regions towards the galactic bulge. The stars in
 the first of these two regions were as close as
 5$\degr$ from the Galactic Center and exhibited
 very significant interstellar extinction. The
 second region was further away and less extincted.
 While most of the sources had colors that 
 could be reconciled with reddened
dwarf or giant late-type stars, some objects 
appeared to have outstanding colors. In general,
the majority of the sources in our sample have
no optical counterpart and very few have spectral
determinations or bibliographical entries.

We detected 20 new binary and 4 new triple systems.
The projected separations are as short as 7\,mas
and the brightness ratios as high as $\Delta$K=6.5\,mag.
Some of these systems could be followed up in principle
by other high angular resolution techniques but
many others would require a combination of
sensitivity and angular resolution which can
only be obtained at present by LO.
We also detect two sources which appear
to be resolved on the milliarcsecond scale, one
of them being a OH-IR star, and the other being
in a pair which shows a spectral energy
distribution rising in the infrared.

We have evaluated the probability of finding
chance associations from the average cumulative
stellar density in the two regions of our
sample, and we found that for the given separations
and magnitude differences the probability
is extremely low.

With this paper we have increased the number
of unresolved stars with reliable upper
limits in the milliarcsecond range, observed
in a homogeneous way with the same instrument
and method, to about 350. This list, to be further
extended with more observations currently under
analysis, can serve as a useful
reference catalogue for interferometric calibrators.

\begin{acknowledgements}
AR acknowledges support
from the
ESO Director General's Discretionary Fund.
OF acknowledges financial support from MICINN through 
a {\it Juan de la Cierva} fellowship and from 
\emph{MCYT-SEPCYT Plan Nacional I+D+I AYA\#2008-01225}.
We are grateful to the Paranal Observatory staff
and in particular to Dr. E. Mason.
This research made use of the Simbad database,
operated at the CDS, Strasbourg, France, and
of data products from the Two Micron All Sky Survey, 
which is a joint project of the University of Massachusetts 
and the Infrared Processing and Analysis Center/California Institute 
of Technology, funded by the National Aeronautics and 
Space Administration and the National Science Foundation.
\end{acknowledgements}

\longtab{2}{
\begin{longtable}{lccrrrll}
\caption{\label{lo_complete}
List of the recorded occultation events}\\
\hline\hline
\multicolumn{1}{c}{2MASS id}&
\multicolumn{1}{c}{Date}&
\multicolumn{1}{c}{UT}&
\multicolumn{1}{c}{J}&
\multicolumn{1}{c}{H}&
\multicolumn{1}{c}{K}&
\multicolumn{1}{c}{Sp}&
\multicolumn{1}{c}{Cross-Id} \\
\hline
\endfirsthead
\caption{continued.}\\
\hline\hline
\multicolumn{1}{c}{2MASS id}&
\multicolumn{1}{c}{Date}&
\multicolumn{1}{r}{UT}&
\multicolumn{1}{c}{J}&
\multicolumn{1}{c}{H}&
\multicolumn{1}{c}{K}&
\multicolumn{1}{c}{Sp}&
\multicolumn{1}{c}{Cross-Id} \\
\hline
\endhead
\hline
\endfoot
18012777-2538123 & 25-09-09 & 23:14:10 & 8.35 & 7.10 & 6.56 & & \\
18014156-2539527 & 25-09-09 & 23:19:44 & 9.41 & 7.82 & 6.93 & & \\
18015800-2549077 & 25-09-09 & 23:25:04 & 9.42 & 7.53 & 6.61 & & \\
18020084-2543076 & 25-09-09 & 23:28:49 & 9.51 & 7.85 & 6.85 & & \\
18014561-2536014 & 25-09-09 & 23:31:22 & 10.41 & 8.32 & 7.08 & & \\
18020469-2553488 & 25-09-09 & 23:35:55 & 8.89 & 7.57 & 7.01 & & \\
18015392-2534372 & 25-09-09 & 23:40:36 & 9.27 & 7.55 & 6.70 & & \\
18023182-2539066 & 25-09-09 & 23:54:50 & 7.44 & 5.94 & 5.22 & & \\
18024042-2549567 & 25-09-09 & 23:59:45 & 9.74 & 8.01 & 7.15 & & \\
18022636-2555373 & 26-09-09 & 0:01:47 & 6.97 & 6.24 & 5.98 & K5 & HD 314951 \\
18024090-2552007 & 26-09-09 & 0:03:42 & 12.45 & 11.10 & 8.90 & & \\
18015464-2529273 & 26-09-09 & 0:05:36 & 9.39 & 7.35 & 6.32 & & \\
18024492-2535538 & 26-09-09 & 0:08:21 & 8.62 & 7.52 & 6.94 & & \\
18022315-2530237 & 26-09-09 & 0:11:33 & 10.12 & 8.16 & 7.08 & & \\
18030392-2542136 & 26-09-09 & 0:14:25 & 9.65 & 7.97 & 7.16 & & \\
18030699-2543447 & 26-09-09 & 0:16:26 & 9.18 & 7.51 & 6.77 & & \\
18030504-2536574 & 26-09-09 & 0:19:25 & 8.18 & 6.64 & 5.84 & & \\
18025652-2532392 & 26-09-09 & 0:22:03 & 7.35 & 6.15 & 5.60 & & \\
18030887-2535164-1 & 26-09-09 & 0:24:15 & 8.74 & 7.39 & 7.13 & & \\
18030887-2535164-2 & & & & & 8.28 & & \\
18030533-2532570 & 26-09-09 & 0:26:25 & 6.77 & 5.46 & 4.90 & & \\
18032159-2538369 & 26-09-09 & 0:28:31 & 6.87 & 5.73 & 5.26 & & IRAS 18002-2538 \\
18032661-2538323 & 26-09-09 & 0:31:54 & 8.97 & 7.61 & 7.08 & & \\
18033221-2541506 & 26-09-09 & 0:34:32 & 12.80 & 9.20 & 7.40 & & \\
18033459-2544130 & 26-09-09 & 0:36:54 & 9.29 & 7.79 & 7.14 & & \\
18033362-2547375 & 26-09-09 & 0:39:47 & 8.17 & 6.61 & 5.66 & & IRAS 18004-2547 \\
18030885-2527400 & 26-09-09 & 0:43:18 & 6.65 & 5.41 & 4.88 & & \\
18034631-2533153 & 26-09-09 & 0:49:56 & 7.29 & 5.99 & 5.28 & & IRAS 18006-2533 \\
18035693-2539570 & 26-09-09 & 0:51:48 & 7.32 & 5.96 & 5.33 & & \\
18035200-2547316 & 26-09-09 & 0:54:43 & 10.25 & 8.52 & 7.62 & & \\
18032656-2526103 & 26-09-09 & 0:56:37 & 11.02 & 8.71 & 7.18 & & \\
18040728-2539532 & 26-09-09 & 0:58:54 & 7.08 & 5.65 & 4.86 & & \\
18033401-2525313 & 26-09-09 & 1:02:11 & 8.84 & 7.48 & 6.84 & & \\
18041209-2544257 & 26-09-09 & 1:05:19 &  &  & 9.41 & & \\
18041207-2544225 & 26-09-09 & 1:05:24 & 8.63 & 6.66 & 5.59 & & IRAS 18010-2544  \\
18042112-2533526 & 26-09-09 & 1:10:12 & 7.99 & 6.71 & 6.28 & & \\
18033522-2523185 & 26-09-09 & 1:12:42 & 8.40 & 7.04 & 6.43 & & \\
18043026-2541466 & 26-09-09 & 1:15:59 & 8.55 & 7.01 & 6.29 & & \\
18043001-2531141 & 26-09-09 & 1:18:06 & 7.92 & 6.45 & 5.68 & & \\
18041530-2525329 & 26-09-09 & 1:20:33 & 5.42 & 4.62 & 4.24 & K5 & HD 315063 \\
18040356-2523142 & 26-09-09 & 1:22:36 & 9.01 & 7.59 & 6.93 & & \\
18043132-2527356 & 26-09-09 & 1:24:10 & 11.31 & 8.51 & 7.64 & & \\
18044848-2536204 & 26-09-09 & 1:26:16 & 5.99 & 5.70 & 5.63 & F2/F3II & HD 164972 \\
18044287-2544318 & 26-09-09 & 1:29:56 & 7.68 & 6.41 & 5.82 & & \\
18044594-2527198 & 26-09-09 & 1:32:17 & 9.60 & 8.13 & 7.41 & & \\
18045257-2528135 & 26-09-09 & 1:34:29 & 10.22 & 8.25 & 7.19 & & \\
18045371-2524226 & 26-09-09 & 1:41:35 & 6.19 & 4.94 & 4.37 & & IRAS 18018-2524 \\
18051090-2529257 & 26-09-09 & 1:43:14 & 9.75 & 8.14 & 7.18 & & \\
18050733-2543282 & 26-09-09 & 1:47:21 & 8.42 & 7.24 & 6.71 & & \\
18050977-2543351-1 & 26-09-09 & 1:49:47 & 9.70 & 8.24 & 7.36 & & \\
18050977-2543351-2 & & & & &11.30 & & \\
18052039-2524573 & 26-09-09 & 1:53:44 & 9.83 & 8.13 & 6.91 & & \\
18053347-2537524 & 26-09-09 & 1:56:51 & 8.70 & 7.49 & 6.85 & & \\
18052602-2523384 & 26-09-09 & 1:58:38 & 8.17 & 7.00 & 6.57 & & \\
18053438-2540309 & 26-09-09 & 2:01:33 & 8.73 & 7.53 & 6.87 & & \\
18054382-2537288 & 26-09-09 & 2:03:33 & 8.86 & 7.63 & 6.95 & & \\
18052646-2520179 & 26-09-09 & 2:05:31 & 9.98 & 8.38 & 7.56 & & \\
18054490-2540399 & 26-09-09 & 2:10:07 & 4.95 & 3.67 & 3.15 & & \\
18060176-2526099 & 26-09-09 & 2:13:53 & 4.28 & 3.13 & 2.77 & M6 & IRC -30355 \\
18055576-2540082 & 26-09-09 & 2:17:18 & 5.93 & 4.96 & 4.42 & & \\
18054212-2543196 & 26-09-09 & 2:20:54 & 6.77 & 5.59 & 4.94 & & IRAS 18026-2543 \\
18061707-2523008 & 26-09-09 & 2:24:46 & 8.36 & 7.08 & 6.50 & & \\
18062439-2525353-1 & 26-09-09 & 2:26:21 & 11.43 & 9.00 &10.14 & & \\
18062439-2525353-2 & 26-09-09 & 2:26:21 & 11.43 & 9.00 & 7.33 & & \\
18054372-2514234 & 26-09-09 & 2:30:17 & 4.72 & 3.36 & 2.74 & M8+ & \\
18063774-2529277 & 26-09-09 & 2:32:49 & 14.79 & 12.22 & 8.97 & & IRAS 18035-2529 \\
18061354-2515305 & 26-09-09 & 2:36:23 & 8.74 & 7.41 & 6.78 & & \\
18065092-2529560 & 26-09-09 & 2:40:40 & 5.97 & 5.06 & 4.75 & M & HD 315192 \\
18065490-2530316 & 26-09-09 & 2:43:21 & 9.01 & 7.71 & 7.07 & & \\
18063720-2514416 & 26-09-09 & 2:47:27 & 12.09 & 8.09 & 8.19 & & \\
18064787-2514557 & 26-09-09 & 2:51:10 & 9.52 & 7.85 & 6.85 & & \\
18064977-2537355 & 26-09-09 & 2:53:14 & 8.29 & 7.10 & 6.53 & & \\
18071918-2524448 & 26-09-09 & 2:55:55 & 7.54 & 6.54 & 6.16 & & \\
18071398-2516474 & 26-09-09 & 2:59:12 & 8.46 & 8.19 & 7.91 & & HD 315178 \\
18072061-2517190 & 26-09-09 & 3:01:32 & 9.18 & 6.87 & 5.38 & & \\
18063098-2509319 & 26-09-09 & 3:04:08 & 7.33 & 6.29 & 5.87 & & IRAS 18034-2509 \\
18072731-2532426 & 26-09-09 & 3:06:38 & 8.02 & 6.85 & 6.23 & & \\
18074347-2524288 & 26-09-09 & 3:08:58 & 7.02 & 5.78 & 5.14 & & \\
18071147-2509208 & 26-09-09 & 3:13:48 & 9.55 & 8.09 & 7.35 & & \\
18073464-2534117 & 26-09-09 & 3:16:25 & 6.44 & 5.94 & 5.83 & G7III+... & HD 165530 \\
18074951-2532392 & 26-09-09 & 3:22:19 & 8.01 & 6.60 & 5.78 & & IRAS 18047-2533 \\
18081328-2523474 & 26-09-09 & 3:24:52 & 6.77 & 5.70 & 5.21 & & \\
18081825-2521370 & 26-09-09 & 3:27:06 & 8.74 & 7.60 & 7.01 & & \\
18072725-2506165 & 26-09-09 & 3:29:38 & 7.15 & 6.97 & 6.86 & BIa & HD 165517 \\
18082752-2518296 & 26-09-09 & 3:32:12 & 10.66 & 8.59 & 7.32 & & \\
18083425-2523306 & 26-09-09 & 3:36:04 & 11.66 & 9.31 & 7.76 & & \\
18543875-2403222 & 26-09-09 & 23:23:26 & 6.95 & 6.06 & 5.72 & & CD-24 14823 \\
18550100-2408394 & 26-09-09 & 23:30:15 & 9.36 & 8.38 & 8.07 & & \\
18541928-2358016 & 26-09-09 & 23:32:31 & 10.51 & 9.65 & 9.35 & & \\
18551118-2413072 & 26-09-09 & 23:36:52 & 8.83 & 7.86 & 7.39 & & \\
18550780-2403564 & 26-09-09 & 23:41:25 & 7.44 & 6.54 & 6.23 & & \\
18552223-2412461 & 26-09-09 & 23:46:11 & 10.37 & 9.80 & 9.66 & & \\
18550997-2401055 & 26-09-09 & 23:48:36 & 7.37 & 6.79 & 6.58 & K1III & HD 175159 \\
18545320-2357035-1 & 26-09-09 & 23:51:43 & 8.54 & 8.01 & 7.89 & & TYC 6860-1491-1 \\
18545320-2357035-2 & & & & &11.07 & & TYC 6860-1491-1 \\
18553233-2412387 & 26-09-09 & 23:54:53 & 7.94 & 6.91 & 6.44 & & \\
18553706-2411230 & 26-09-09 & 23:58:20 & 10.69 & 9.76 & 9.54 & & \\
18550303-2355559 & 27-09-09 & 0:00:52 & 8.87 & 8.06 & 7.81 & & \\
18554320-2411418 & 27-09-09 & 0:03:48 & 10.03 & 9.41 & 9.19 & & \\
18553293-2359253 & 27-09-09 & 0:06:37 & 9.02 & 8.09 & 7.72 & & \\
18554220-2415597 & 27-09-09 & 0:09:11 & 9.65 & 8.76 & 8.32 & & \\
18550756-2354134 & 27-09-09 & 0:10:21 & 8.55 & 7.82 & 7.65 & & \\
18554874-2401474 & 27-09-09 & 0:12:40 & 9.07 & 8.15 & 7.74 & & \\
18554945-2400538 & 27-09-09 & 0:14:28 & 10.20 & 9.56 & 9.42 & & \\
18555913-2403339 & 27-09-09 & 0:17:56 & 10.04 & 9.17 & 8.96 & & \\
18555632-2400073 & 27-09-09 & 0:20:14 & 11.20 & 10.46 & 10.30 & & \\
18561680-2406431 & 27-09-09 & 0:30:12 & 10.40 & 9.66 & 9.46 & & \\
18555737-2354145 & 27-09-09 & 0:33:27 & 8.37 & 7.44 & 7.14 & & \\
18561788-2358243 & 27-09-09 & 0:36:56 & 11.31 & 10.52 & 10.39 & & \\
18562006-2357173 & 27-09-09 & 0:40:04 & 11.15 & 10.31 & 10.15 & & \\
18553777-2349547 & 27-09-09 & 0:42:05 & 10.71 & 9.86 & 9.53 & & \\
18552401-2349283 & 27-09-09 & 0:43:57 & 8.79 & 7.84 & 7.56 & & \\
18563747-2408447 & 27-09-09 & 0:49:24 & 10.73 & 9.88 & 9.65 & & \\
18563774-2410374 & 27-09-09 & 0:53:15 & 8.95 & 8.04 & 7.67 & & \\
18564920-2400465 & 27-09-09 & 0:56:10 & 10.99 & 10.21 & 10.02 & & \\
18564563-2354023 & 27-09-09 & 1:01:11 & 10.98 & 10.09 & 9.86 & & \\
18565387-2355147 & 27-09-09 & 1:04:24 & 8.88 & 8.14 & 7.95 & & \\
18570282-2358475 & 27-09-09 & 1:06:57 & 9.23 & 8.28 & 8.03 & & \\
18564695-2349245 & 27-09-09 & 1:11:43 & 10.90 & 10.24 & 10.06 & & \\
18561672-2345402 & 27-09-09 & 1:14:04 & 11.40 & 10.61 & 10.40 & & \\
18570432-2351321 & 27-09-09 & 1:16:26 & 10.15 & 9.21 & 8.96 & & \\
18570946-2352076 & 27-09-09 & 1:18:26 & 9.57 & 9.40 & 9.32 & & CD-24 14860B \\
18570922-2351582 & 27-09-09 & 1:18:33 & 9.42 & 9.28 & 9.21 & A1V+... & HD 175601 \\
18570262-2410283 & 27-09-09 & 1:21:48 & 8.92 & 8.21 & 7.98 & & \\
18565615-2346249 & 27-09-09 & 1:24:19 & 10.71 & 10.09 & 10.03 & & \\
18572854-2402401 & 27-09-09 & 1:27:10 & 6.51 & 5.57 & 5.18 & & \\
18565536-2344580 & 27-09-09 & 1:28:41 & 8.98 & 8.14 & 7.74 & & \\
18572506-2348190 & 27-09-09 & 1:33:49 & 5.92 & 5.09 & 4.60 & & \\
18574519-2354457 & 27-09-09 & 1:38:20 & 9.74 & 8.87 & 8.65 & & \\
18573657-2405454 & 27-09-09 & 1:40:56 & 9.65 & 9.03 & 8.90 & & \\
18574575-2349490 & 27-09-09 & 1:43:03 & 8.77 & 7.97 & 7.86 & & \\
18575364-2400373 & 27-09-09 & 1:46:05 & 8.92 & 8.11 & 7.79 & & \\
18573272-2343187 & 27-09-09 & 1:48:57 & 10.89 & 10.01 & 9.86 & & \\
18575391-2345582 & 27-09-09 & 1:53:22 & 8.96 & 8.01 & 7.73 & & \\
18580998-2355311 & 27-09-09 & 1:55:33 & 10.56 & 9.98 & 9.85 & & \\
18580488-2403296 & 27-09-09 & 2:03:58 & 10.91 & 10.10 & 9.83 & & \\
18580978-2343262 & 27-09-09 & 2:06:01 & 9.61 & 8.65 & 8.35 & & \\
18581661-2343555 & 27-09-09 & 2:08:40 & 9.35 & 9.18 & 9.09 & F2V & HD 175851 \\
18580940-2340445 & 27-09-09 & 2:11:24 & 6.48 & 5.53 & 5.22 & & IRAS 18551-2344 \\
18581647-2340114 & 27-09-09 & 2:15:48 & 8.22 & 7.34 & 6.89 & & \\
18580265-2337079-1 & 27-09-09 & 2:18:59 & 9.08 & 8.52 & 9.37 & & \\
18580265-2337079-2 & & & & & 9.06 & & \\
18582165-2338403 & 27-09-09 & 2:21:34 & 8.87 & 8.32 & 8.21 & & \\
18585453-2349519 & 27-09-09 & 2:26:02 & 6.49 & 5.51 & 5.07 & & \\
18585782-2351145 & 27-09-09 & 2:28:29 & 10.65 & 9.91 & 9.79 & & \\
18585736-2342464 & 27-09-09 & 2:31:53 & 8.59 & 7.74 & 7.55 & & \\
18581996-2334007 & 27-09-09 & 2:35:20 & 10.48 & 9.91 & 9.80 & & \\
18591262-2346443 & 27-09-09 & 2:38:07 & 11.69 & 11.01 & 10.86 & & \\
18591792-2350546 & 27-09-09 & 2:42:38 & 10.11 & 9.50 & 9.37 & & \\
18583937-2332312 & 27-09-09 & 2:45:54 & 9.14 & 8.34 & 8.12 & & \\
18592407-2341426 & 27-09-09 & 2:47:34 & 8.69 & 7.86 & 7.66 & & \\
18593674-2346506 & 27-09-09 & 2:53:35 & 11.12 & 10.31 & 10.08 & & \\
18592411-2335094 & 27-09-09 & 2:55:58 & 10.46 & 9.83 & 9.78 & & \\
18594214-2342263 & 27-09-09 & 2:57:23 & 8.41 & 7.95 & 7.68 & G4 & AR Sgr \\
18585553-2329395 & 27-09-09 & 3:00:45 & 10.12 & 9.50 & 9.39 & & \\
18585480-2329085 & 27-09-09 & 3:02:56 & 9.75 & 9.23 & 9.12 & & \\
18593194-2355053 & 27-09-09 & 3:05:03 & 9.39 & 8.48 & 7.97 & & \\
18595823-2339176 & 27-09-09 & 3:08:09 & 9.50 & 8.83 & 8.65 & & \\
18595396-2334060 & 27-09-09 & 3:11:17 & 10.78 & 9.84 & 9.61 & & \\
19000813-2343368 & 27-09-09 & 3:13:02 & 11.61 & 10.88 & 10.71 & & \\
19001313-2340112 & 27-09-09 & 3:16:07 & 7.63 & 6.66 & 6.27 & & \\
18595891-2331159 & 27-09-09 & 3:17:55 & 9.50 & 8.57 & 8.29 & & \\
19001505-2348245 & 27-09-09 & 3:22:45 & 10.17 & 9.50 & 9.30 & & \\
19002907-2340467 & 27-09-09 & 3:25:28 & 9.17 & 8.58 & 8.45 & & \\
19001602-2349574 & 27-09-09 & 3:28:15 & 8.47 & 7.61 & 7.27 & & \\
19000444-2326074 & 27-09-09 & 3:30:58 & 9.39 & 8.81 & 8.60 & & \\
19004442-2339386 & 27-09-09 & 3:34:32 & 7.88 & 7.00 & 6.60 & & \\
19003871-2330374 & 27-09-09 & 3:36:30 & 10.09 & 9.48 & 9.34 & & \\
19005121-2333311 & 27-09-09 & 3:39:53 & 10.82 & 10.17 & 10.03 & & \\
19004262-2326331 & 27-09-09 & 3:44:04 & 8.18 & 7.60 & 7.43 & & CD-23 14943 \\
19002919-2323136 & 27-09-09 & 3:46:23 & 10.93 & 10.27 & 10.08 & & \\
19004796-2324055 & 27-09-09 & 3:50:42 & 8.88 & 8.10 & 7.93 & & \\
19011673-2332578 & 27-09-09 & 3:53:21 & 10.06 & 9.13 & 8.86 & & \\
19012148-2332350 & 27-09-09 & 3:55:59 & 10.55 & 9.89 & 9.73 & & \\
19012789-2336241 & 27-09-09 & 3:59:28 & 8.84 & 8.02 & 7.72 & & \\
19011187-2323074 & 27-09-09 & 4:01:26 & 10.21 & 9.63 & 9.49 & & \\
19012308-2342339 & 27-09-09 & 4:03:49 & 9.14 & 8.18 & 7.81 & & \\
19014063-2332313 & 27-09-09 & 4:06:08 & 9.61 & 9.03 & 8.88 & & \\
19005076-2317221 & 27-09-09 & 4:11:11 & 7.37 & 6.80 & 6.69 & G8IV & HD 176364 \\
19010885-2317367 & 27-09-09 & 4:13:26 & 10.12 & 9.85 & 9.74 & & \\
19015719-2335001 & 27-09-09 & 4:16:07 & 9.14 & 8.18 & 7.87 & & \\
19015242-2339279 & 27-09-09 & 4:18:56 & 8.66 & 7.94 & 7.71 & & \\
19013557-2316469 & 27-09-09 & 4:22:49 & 9.59 & 9.34 & 9.25 & & \\
19021893-2330540 & 27-09-09 & 4:26:44 & 9.51 & 8.85 & 8.61 & & \\
\end{longtable}
}
\end{document}